\documentclass[aps,prx,a4paper,twocolumn,longbibliography,floatfix,superscriptaddress]{revtex4-2}

\usepackage{amssymb,amsmath,amsfonts}
\usepackage{graphicx}
\usepackage[dvipsnames]{xcolor}
\usepackage[bookmarksopen,colorlinks,linkcolor=blue,citecolor=olive,urlcolor=red]{hyperref}
\usepackage{bm}

\def\div{\mathop{\rm div}}

\def\br{\mathbf{r}}
\def\bj{\mathbf{j}}

\def\vp{\varphi}
\def\eps{\varepsilon}

\newcommand{\corr}[1]{\langle#1\rangle}

\def\be{\begin{equation}}
\def\ee{\end{equation}}

\begin{document}

\title{Thermal phase slips in superconducting films}

\author{Mikhail A.\ Skvortsov}
\affiliation{L.~D.\ Landau Institute for Theoretical Physics,  Chernogolovka 142432, Russia}

\author{Artem V.\ Polkin}
\affiliation{L.~D.\ Landau Institute for Theoretical Physics,  Chernogolovka 142432, Russia}

\affiliation{Laboratory for Condensed Matter Physics, HSE University, Moscow 101000, Russia}

\date{\today}

\begin{abstract}
A dissipationless supercurrent state in superconductors can be destroyed by thermal fluctuations. Thermally activated phase slips provide a finite resistance of the sample and are responsible for dark counts in superconducting single photon detectors. The activation barrier for a phase slip is determined by a space-dependent saddle-point (instanton) configuration of the order parameter. In the one-dimensional wire geometry, such a saddle point has been analytically obtained by Langer and Ambegaokar in the vicinity of the critical temperature, $T_c$, and for arbitrary bias currents below the critical current $I_c$. In the two-dimensional geometry of a superconducting strip, which is relevant for photon detection, the situation is much more complicated. Depending on the ratio $I/I_c$, several types of saddle-point configurations have been proposed, with their energies being obtained numerically. We demonstrate that the saddle-point configuration for an infinite superconducting film at $I\to I_c$ is described by the exactly integrable Boussinesq equation solved by Hirota's method. The instanton size is $L_x\sim\xi(1-I/I_c)^{-1/4}$ along the current and $L_y\sim\xi(1-I/I_c)^{-1/2}$ perpendicular to the current, where $\xi$ is the Ginzburg-Landau coherence length. The activation energy for thermal phase slips scales as $\Delta F^\text{2D}\propto (1-I/I_c)^{3/4}$. For sufficiently wide strips of width $w\gg L_y$, a half-instanton is formed near the boundary, with the activation energy being 1/2 of $\Delta F^\text{2D}$.
\end{abstract}

\maketitle

Superconductivity is the ability to conduct electric current without dissipation due to the coherent flow of the Cooper pairs condensate. 
However such an idealized picture is undermined by vortex motion and spontaneous phase slips, either thermal \cite{LA,McCumber} or quantum \cite{QPS-Zaikin,QPS-Zhao}. Phase slips drive a superconductor out of its dissipationless supercurrent state, causing finite voltage and therefore dissipation. When quantum phase slips are coherent, they destroy phase integrity of Cooper pairs and mediate the superconductor-to-insulator transition \cite{Bezryadin2000,Mooij,Astafiev2012,Astafiev2022}.

The most convenient platform for studying individual thermal phase slips is provided by superconducting nanowire single-photon detectors (SNSPDs) \cite{Goltsman01,Natarajan12,Zadeh2021}. These devices are engineered to rapidly restore a disrupted supercurrent state, preventing the formation of the resistive state when phase-slip proliferation creates a complex dynamic with intermittent regions of superconducting and normal phases \cite{IvlevKopnin84,WeberKramer1991}. In SNSPDs, phase slips trigger \emph{dark counts}, when a voltage pulse arises in the absence of an incoming photon \cite{Engel2006,Kitaygorsky2007}. While detrimental to photon-counting fidelity, these events enable direct measurement of the phase slip rate \cite{Bartolf2010,Semenov20}.

The rate of thermal phase slips is determined by the height $\Delta F$ of the potential barrier protecting the metastable supercurrent state and corresponding to a saddle point of the free energy. Once the latter is reached, superconductivity is (locally) destroyed, producing a finite voltage. Although the emerging dynamical state can be rather involved \cite{IvlevKopnin84,WeberKramer1991}, to exponential accuracy, 
computing the decay rate $e^{-\Delta F/T}$ of a uniform supercurrent reduces to finding a stationary saddle-point the system must overcome for a phase slip to occur at a later stage.

A current-biased Josephson junction controlled by the phase difference $\vp$ provides the simplest example with the barrier $\Delta F^\text{0D} = U(\vp_\text{max}) - U(\vp_\text{min})$, where $\vp_\text{min}$ and $\vp_\text{max}$ is a minimum and a neighboring maximum of the tilted washboard potential $U(\vp)=-E_J\cos\vp-(\hbar/2e)I\vp$, and $E_J$ is the Josephson energy.

The theory of thermal phase slips in thin superconducting wires in the vicinity of the critical temperature, $T_c$, has been developed by Langer and Ambegaokar (LA) long time ago \cite{LA,McCumber}. They have analytically identified the localized saddle-point (instanton) solution for the Ginzburg-Landau (GL) equations, with the suppression of the modulus of the order parameter $|\Delta|$ at the center of the instanton being accompanied by the increase of the phase gradient. The instanton size $L_x\sim\xi(1-I/I_c)^{-1/4}$ diverges when the current $I$ approaches the critical current $I_c$ [$\xi(T)$ is the GL coherence length], resulting in the power-law scaling of the barrier:
\be
\label{F-LA}
  \Delta F^\text{1D} 
  \approx 
  c_1
  \eps_\text{cond}A\xi (1-I/I_c)^{5/4} ,
\ee
where $\eps_\text{cond}(T) = H_c^2/8\pi$
is the condensation energy density, $A$ is the wire cross-section, 
and $c_1 = 2^{25/4} 3^{-5/4} 5^{-1} = 3.86$.

Yet the one-dimensional (1D) theory of LA fails to describe dark counts in SNSPDs with a superconducting strip of the width $w\gg\xi$ inside, and generalization of the theory to the two-dimensional (2D) case is required. This appears to be a nontrivial task due to (i) (co)existence of different types of saddles: both topologically trivial (LA-like, when the order parameter is suppressed in the center of the nucleus but remains finite) and topologically nontrivial (single vortex, vortex-antivortex pair), (ii) inhomogeneous current distribution, and (iii) restricted geometry. As a result, previous studies were limited to providing approximate or numerical estimates of the instanton configuration and the associated energy barrier.
Inspired by hot-spot physics of nonequilibrium photon detection \cite{Goltsman01}, a popular scenario was to interpret equilibrium dark counts as thermal unbinding of vortex-antivortex pairs (topologically nontrivial saddle point) \cite{Engel2006,Kitaygorsky2007,Bulaevsky}.
The most comprehensive numerical search for the saddle-point solutions in the SNSPD context has been carried out by Vodolazov \cite{Vodolazov2012}, who considered nanowires of different widths $w$ (varied from $4.5\xi$ up to $30\xi$) in the vicinity of $T_c$. He has shown that the topology of the lowest-energy saddle point is controlled by the current: it is vortex-like (vanishing $\Delta$) at low $I$ and topologically trivial (non-vanishing $\Delta$) near $I_c$. 

\begin{figure}
\includegraphics[width=0.95\columnwidth]{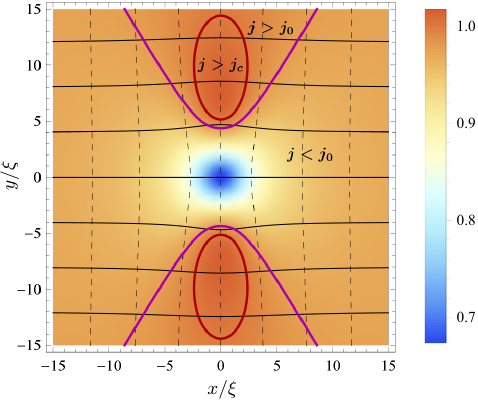}
\caption{Saddle-point configuration of the current $\bj(x,y)$ responsible for the thermal phase slip at $I/I_c=0.98$. The supercurrent follows the solid lines, and its magnitude (normalized by the current density $j_0$ at infinity) is shown by the color scale. Contours of constant phase appear dashed. In the regions encircled by thick solid lines, $j$ exceeds the nominal critical current density $j_c$ for a uniform flow.}
\label{F:tok}
\end{figure}

In this Letter we report on the first \emph{analytical} determination of the phase slip rate in superconducting films. Motivated by recent experiments with micron-wide-strip SNSPDs \cite{SNSPD-wide-Korneeva,SNSPD-wide-Charaev}, which is nearly 1000 times larger than the coherence length, we consider an infinite 2D geometry without edges. 
Working in the GL region close to $T_c$, we obtain the following expression for the activation barrier valid near the critical current:
\be
\label{DF-res}
  \Delta F^\text{2D}
  \approx
  c_2
  \eps_\text{cond} d \xi^2
  (1-I/I_c)^{3/4} ,
\ee
where $d$ is the film thickness, and $c_2 = 2^{27/4} 3^{-9/4} \pi = 28.55$. The current density at the saddle-point solution normalized by its value at infinity is given by anisotropic Lorentz-like functions
\begin{subequations}
\label{j-res}
\begin{gather}
\label{jx-res}
  \frac{j_x}{j_0}
  = 
  1 
  -
  \frac{18 \eps^2 \xi^2 ( 2\eps x^2 - 3 \eps^2 y^2 + 3 \xi^2 )}
  {(2 \eps x^2 + 3 \eps^2 y^2 + 3 \xi^2 )^2} ,
\\
  \frac{j_y}{j_0}
  = 
  - 
  \frac{72 \eps^3 \xi^2 x y}
  {(2 \eps x^2 + 3 \eps^2 y^2 + 3 \xi^2)^2} ,
\end{gather}
\end{subequations}
where
$\eps \approx \sqrt{(8/3)(1-I/I_c)} \ll 1$ is the small parameter of the theory. The size of the optimal fluctuation along the applied current ($\bj=j_0\mathbf{e}_x$) coincides with the size of the 1D LA instanton: $L_x\sim\xi/\sqrt\eps\sim\xi(1-I/I_c)^{-1/4}$, whereas the transverse dimension of the instanton is parametrically larger: $L_y\sim\xi/\eps\sim\xi(1-I/I_c)^{-1/2}\gg L_x$
(a similar strong anisotropy at $I\to I_c$ has recently been reported in Ref.\ \cite{Zuev}). 
This explains the scaling \eqref{DF-res} of the 2D barrier:
$F^\text{2D}\sim (L_y/\xi) F^\text{1D}$, changing the 1D exponent of 5/4 to 3/4.

In Fig.\ \ref{F:tok}, we visualize the instanton current pattern at $\eps=0.23$ ($I/I_c=0.98$). The supercurrent flows along black solid lines, and its magnitude $|\bj(\br)|$ is shown by the color scale. The current density has a dip, with a minimum at the center of the instanton.
The obtained solution conserves the net current, and $\bj(\br)$ is redistributed to compensate the suppression near the origin by an excess current ($|\bj|>j_0$) in the regions 
$y^2 > \xi^2/\eps^2 + 2x^2/3\eps$.
Remarkably, the supercurrent exceeds the nominal critical current density $j_c$ in two regions along the $y$ axis encircled by the thick line.
Such an ``overheated'' configuration---forbidden in the uniform case---is stabilized by a finite $|\Delta|$ gradient.

Now we briefly describe the approach for analytical determination of the 2D instanton \eqref{j-res}. The starting point is GL free energy written in the dimensionless form:
\be
\label{GL}
  F = 
  C
  \int dx\,dy
  \left(
    |\nabla\Delta|^2 - |\Delta|^2 + |\Delta|^4/2 
  \right) ,
\ee
where coordinates are measured in units of $\xi$, the order parameter field $\Delta(\br)=|\Delta(\br)|e^{i\vp(\br)}$ is normalized by its equilibrium value without current, and $C=2 \eps_\text{cond} d \xi^2$. 
Assuming the film width $w$ is much shorter than the Pearl length $\Lambda$, we neglect magnetic field effects. This approximation works fairly well even for the widest ($w\sim1\,\mu\text{m}$) available films, with $\Lambda\sim90\,\mu\text{m}$ for NbN \cite{SNSPD-wide-Korneeva} and $600\,\mu\text{m}$ for MoSi \cite{SNSPD-wide-Charaev}.
The current density is $\bj = |\Delta|^2 \nabla\vp$, and we consider configurations approaching a given $j_0\mathbf{e}_x$ at infinity. The standard uniform-current state, providing a free-energy minimum, is characterized by a linear phase growth $\vp_0=Ax$, suppressed order parameter $|\Delta_0|^2=1-A^2$, and $j_0=(1-A^2)A$. The critical current $j_c=2/3^{3/2}$ is achieved at $A_c=1/\sqrt3$ \cite{Tin}.

The instanton configuration corresponds to a spatially localized inhomogeneous saddle point of the functional \eqref{GL} asymptotically approaching $\Delta_0(\br)$ at $r\to\infty$.
To find it, one has to solve a set of coupled, nonlinear 2D partial differential GL equations governing the order parameter’s modulus and phase. 
The main difficulty is that neither $|\Delta(\br)|$ nor $\vp(\br)$ can be excluded to formulate a theory in terms of a single real function. 
To construct such a representation, we suggest to 
write the supercurrent density $\bj(\br)$, which obeys the current conservation law $\div\bj=0$, in terms of a scalar \emph{stream function} $\psi(\br)$:
\be
\label{tok-rot}
  \bj = (\psi_y,-\psi_x) .
\ee
Hereafter subscripts mean derivatives: $\psi_x=\partial_x\psi$, etc. 
The representation \eqref{tok-rot} is a standard trick in the 2D hydrodynamics \cite{Dean1965,Falkovich,Fetter2017}, reducing the number of unknown real functions from two to one.
The stream function is harmonic for an incompressible fluid and not harmonic if $|\Delta|$ varies in space.
A uniform supercurrent flow corresponds to the stream function $\psi_0=j_0y$, which plays the role of the boundary condition for the instanton solution at infinity: $\psi\to j_0y$.

In order to be able to formulate the theory in terms of the stream function, one has to answer the principal question: Can the complex order parameter field $\Delta(\br)$ be unequivocally reconstructed for a given $\psi(\br)$?

First we note that, by definition,
\be
\label{psi-phi}
  \nabla\psi \cdot \nabla\vp = 0 ,
\ee
indicating that $\psi$ and $\vp$ is a pair of orthogonal curvilinear coordinates on the plane.
In order to find $\vp(\br)$ for a given $\psi(\br)$, one has to solve Eq.\ \eqref{psi-phi}. It is a linear first-order partial differential equation that can be solved by the method of characteristics with the boundary condition $\vp\to Ax$ at infinity.
A global solution for $\vp(\br)$ exists provided $\nabla\psi\neq0$ everywhere on the 2D plane.
In this case, the characteristics of Eq.\ \eqref{psi-phi}
(dashed lines in Fig.\ \ref{F:tok}), which go vertically at $y=\pm\infty$, bend near the instanton center without intersecting, thereby uniquely defining the phase everywhere.
Put differently, if the superflow at the instanton solution is laminar and vortex-free ($\bj\neq0$), it can be completely described by the stream function. 

The applicability of the stream-function approach to the phase slip determination at $I\to I_c$ is related to \emph{fragility} of the uniform superconducting state \cite{LA}, when even a tiny perturbation can switch the system to the normal state. This excludes vortex-like configurations due to a finite energy of the core \cite{Vodolazov2012}.
In the absence of vortices, the phase is uniquely determined by $\psi(\br)$. Once $\vp(\br)$ is known, $|\Delta(\br)|$ is immediately obtained from the local expression for the current.
In this way, we arrive at a theory formulated in terms of the stream function only, but the price is that the free energy $\Delta F[\psi]$ is nonlocal (and can be obtained only numerically) and nonlinear.

The program outlined above can be successfully implemented at $I\to I_c$ using perturbation theory. To this end, we write the stream function as
$\psi = j_0 (y + f_y)$, where $f(x,y)$ is an unknown function vanishing at $I=I_c$. Modification of the phase will be described by the function $g(x,y)$:
$\vp = A (x + g)$.
Substituting these expressions into Eq.\ \eqref{psi-phi}, we get an equation for $g$:
\be
\label{fg-eq}
  f_{xy} (1 + g_x) + ( 1 + f_{yy} ) g_y = 0 ,
\ee
which should be solved for a given $f$.
Near the critical current, when $f$ is small, the solution can be expanded as 
$g = g^{(1)} + g^{(2)} + \dots$, where $g^{(n)}\propto f^n$.
In the first order, we get a local relation 
$g^{(1)} = -f_x$, whereas higher order corrections are nonlocal and should be obtained by solving differential equations
$g^{(2)}_y = (f_{xx} + f_{yy}) f_{xy}$, 
$g^{(3)}_y = - (f_{xx} + f_{yy}) f_{xy} f_{yy} - g^{(2)}_x f_{xy}$, etc.
The order parameter modulus is given by
$|\Delta|^2 = |\Delta_0|^2 (1 + f_{yy})/(1 + g_x)$.
Substituting these expressions into Eq.\ \eqref{GL}, we obtain the free energy as a nonlocal functional of the field $f(x,y)$. Its series expansion starts with the quadratic term, which is local and can be brought to the form including the squares of the second ($f_{xx}^2$, $f_{xy}^2$, $f_{yy}^2$) and third ($f_{xxx}^2$, $f_{xxy}^2$, $f_{xyy}^2$, $f_{yyy}^2$) derivatives. The coefficients in front of these terms depend on $A$ and remain finite at $A\to A_c$, except for $f_{xx}^2$, which is proportional to 
$
\eps = 1-3A^2 \approx \sqrt{(8/3)(1-I/I_c)},
$
vanishing at criticality.
This is the reason one has to keep the next-order term $f_{xxx}^2$. Balancing $\eps f_{xx}^2$ and $f_{xxx}^2$ gives the power-law scaling of the instanton size along the current, $L_x\sim\xi/\sqrt\eps$, coinciding with that of the 1D LA instanton. The instanton size perpendicular to the current can be estimated by comparing $\eps f_{xx}^2$ and $f_{xy}^2$ that yields $L_y\sim\xi/\eps$. Thus, approaching $I_c$, the 2D instanton becomes strongly elongated in the direction perpendicular to the current \cite{Zuev}. As a consequence of the above scaling, only the terms $\eps f_{xx}^2$, $f_{xy}^2$ and $f_{xxx}^2$ should be retained among all quadratic terms near the criticality.

The form of the cubic contribution to the free energy ($f^3$) is more complicated, containing a larger number of terms, some being nonlocal. However in the limit $I\to I_c$, the anisotropy of the instanton ($L_y\gg L_x$) suggests that the leading term should have the minimal number of $x$ derivatives and no $y$ derivatives. This requirement selects the term $f_{xx}^3$, which enters locally. As a result, near the criticality the free energy takes the form (all coefficients except for the first one are evaluated at $I_c$)
\be
\label{F1}
  F =
  C
  \int dx\,dy
  \left(
    \frac\eps3 f_{xx}^2 
  + \frac29 f_{xy}^2
  + \frac16 f_{xxx}^2
  + \frac29 f_{xx}^3
  \right) .
\ee

Rescaling the coordinates and the function according to
$x = (2\eps)^{-1/2} \bar x$, 
$y = 3^{-1/2} \eps^{-1} \bar y$,
and $f = \bar f/2$,
we arrive at 
$F 
  =
  2^{1/2} 3^{-3/2} \eps^{3/2}
  SC
$,
that already reveals the $\eps^{3/2}$ scaling of the barrier. 
The coefficient is determined by the instanton solution for the parameter-free functional
\be
\label{S1}
  S = 
  \int d\bar x\, d\bar y
  \left(
    \frac12 \bar f_{\bar x\bar x}^2 
  + \frac12 \bar f_{\bar x\bar y}^2
  + \frac12 \bar f_{\bar x\bar x\bar x}^2
  + \frac13 \bar f_{\bar x\bar x}^3
  \right) .
\ee
Its saddle-point equation $\delta S/\delta \bar{f}=0$ is conveniently written in terms of $u=\bar f_{\bar x\bar x}$:
\be
\label{Boussinesq-w}
  u_{\bar x\bar x} + u_{\bar y\bar y} - u_{\bar x\bar x\bar x\bar x} + (u^2)_{\bar x\bar x} = 0 ,
\ee
which is the elliptic form of the Boussinesq equation \cite{Boussinesq1872,BZ2002}. In its original (hyperbolic) formulation, with $\bar y=it$ and $t$ being time, Eq.\ \eqref{Boussinesq-w} describes propagation of shallow water waves in the long-wavelength limit. 
The Boussinesq equation belongs to a celebrated class of nonlinear partial differential equations integrable by the inverse scattering method \cite{inverseproblem}.
A hallmark of the wave-form Boussinesq equation's integrability is the existence of solitary waves (solitons).

In the elliptic case considered, we are interested in a nontrivial localized solution to Eq.\ \eqref{Boussinesq-w} that vanishes at infinity (instanton). 
Such an exact solution can be obtained in the Hirota form \cite{Hirota1973,Hirota1977} and is given by
\be
\label{u-sol}
  u_\text{B} = - 6 \partial_{\bar x}^2\ln(\bar x^2+\bar y^2+3) .
\ee
Hence, $\bar f_\text{B}=- 6 \ln(\bar x^2+\bar y^2+3)$, the dimensionless instanton action \eqref{S1} evaluates to $S=8\pi$, and we arrive at Eq.~\eqref{DF-res} for the value of the activation barrier.

\begin{table}
\caption{
Exact instanton solution for the Boussinesq equation \eqref{Boussinesq-w} and several variational approximations.} \label{T:variational}
\begin{ruledtabular}
\begin{tabular}{cccc}
Ansatz for $\bar f(\bar x,\bar y)$ & $c_1^\text{opt}$ & $c_2^\text{opt}$ & $S^\text{opt}/8\pi$ \\
\hline
$- 6\ln(\bar x^2+\bar y^2+3)$ & & & 1 \\
\hline
$a/(1+c_1^2\bar x^2+c_2^2\bar y^2)^{1/2}$ & 0.4 & 0.4 & 1.120 \\
$a/(1+c_1^2\bar x^2+c_2^2\bar y^2)$ & 0.322 & 0.322 & 1.286 \\
$a/(1+c_1^2\bar x^2)(1+c_2^2\bar y^2)$ & 0.298 & 0.422 & 1.612 \\
$a/\cosh(c_1\bar x)\cosh(c_2\bar y)$ & $0.459$ & 0.543 & 1.616 \\
$a \exp(-c_1^2\bar x^2 - c_2^2\bar y^2)$ & $0.365$ & $0.365$ & 3.003 \\
\end{tabular}
\end{ruledtabular}
\end{table}

While Eq.\ \eqref{u-sol} provides an exact instanton solution for the Boussinesq equation \eqref{Boussinesq-w}, it is instructive to construct a saddle point for the action \eqref{S1} following a variational approach. 
To this end, we restrict ourselves to a number of probe functions $\bar f=az(\bar x,\bar y)$ listed in Table \ref{T:variational}. Each ansatz is characterized by an amplitude $a$ and a two-parametric function $z(\bar x,\bar y)$, with $c_1$ and $c_2$ controlling its decay at large $\bar x$ and $\bar y$, respectively.
First we determine the optimal value of the amplitude:
$a^\text{opt} = - \corr{z_{\bar x\bar x}^2 + z_{\bar x\bar x\bar x}^2 + z_{\bar x\bar y}^2}/\corr{z_{\bar x\bar x}^3}$, where $\corr{\cdots} = \int d\bar x\,d\bar y\,(\cdots)$.
Substituting it back to Eq.\ \eqref{S1} and minimizing with respect to $c_1$ and $c_2$, we obtain the following variational estimate for the instanton action:
\be
\label{S-opt}
  S^\text{opt}[z] = \min\limits_{c_1,c_2}
  \frac{\corr{z_{\bar x\bar x}^2 + z_{\bar x\bar x\bar x}^2 + z_{\bar x\bar y}^2}^3}{6 \corr{z_{\bar x\bar x}^3}^2} .
\ee
The values of $S^\text{opt}$ are listed in the last column of Table~\ref{T:variational}. 
The lower boundedness of the right-hand side of Eq.~\eqref{S-opt} provides an independent evidence for the existence of an instanton solution for the Boussinesq equation. We also see that (i) $S^\text{opt}>8\pi$ for all probe functions, (ii) power-law functions with a small exponent provide a better approximation, and (iii) the best candidates depend on $\bar x^2+\bar y^2$. Collectively, these results confirm Eq.\ \eqref{u-sol} as the saddle-point solution with the minimal energy. 

\begin{figure}[t]
\includegraphics[width=\columnwidth]{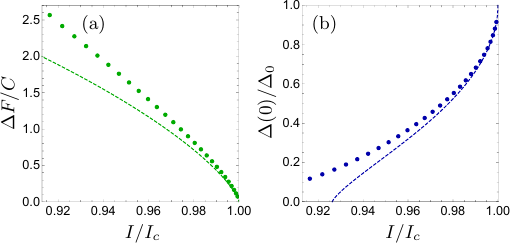}
\caption{Numerical results for the variational Boussinesq-like ansatz: (a) energy barrier (dots) and its asymptotic behavior \eqref{DF-res} (dashed line); (b) order parameter at the center of the instanton (dots) and $\Delta_\text{B}(0)$ (dashed line) normalized by its value at infinity.}
\label{F:Delta-j}
\end{figure}

The current density, $\bj_\text{B}$, for the Boussinesq instanton with $f_\text{B}(x,y)=-3\ln(1+2\eps x^2/3+\eps^2y^2)$ is given by Eq.\ \eqref{j-res}.
It reaches its minimum value at the center, where $j_B(0)/j_0=1-6\eps^2$. 
The order parameter $\Delta_\text{B}(\br)$ corresponding to $f_\text{B}$ also has a minimum at $\br=0$ shown by the dashed line in Fig.\ \ref{F:Delta-j}(b).
Both $j_\text{B}(0)$ and $\Delta_\text{B}(0)$ turn to zero at $\eps=1/\sqrt6$ that corresponds to $I/I_c=0.926$, which is numerically close to the critical current. 
Since the theory assumes that the saddle point is a small modification of the uniform solution, one expects it remains valid in a narrow 5-percent window below $I_c$.

In order to get insight on the behavior of the instanton at larger deviations of $I$ from $I_c$, when the perturbation of $\bj(\br)$ is not small but is still vortex-free, we have performed a numerical variational search of the saddle point. Assuming a Boussinesq-like ansatz for the stream function,
$f(x,y)=-a\ln(1+c_1^2x^2+c_2^2y^2)$, specified by the parameters $a$, $c_1$ and $c_2$ [in the limit $I\to I_c$, given by $f_\text{B}(x,y)$], we reconstruct the phase by numerically solving Eq.\ \eqref{fg-eq} for $g(x,y)$. The free energy is then numerically maximized with respect to $a$ and minimized with respect to $c_{1,2}$. The barrier obtained in this way is shown in Fig.\ \ref{F:Delta-j}(a), with the dashed line illustrating the asymptotic expression \eqref{DF-res}. The order parameter at the instanton center is presented in Fig.\ \ref{F:Delta-j}(b).
Suppression of $\Delta(0)$ nearly follows $\Delta_\text{B}(0)$ (dashed line) up to $0.95\,I_c$, becoming weaker for smaller currents. 
The last obtained point with $\Delta(0)\approx 0.1\Delta_0$ is calculated at $I_1\approx 0.915\,I_c$, and our numerical scheme cannot identify a saddle-point solution below $I_1$.

Given that the order parameter at the center has already lost 90\% of $\Delta_0$ at the last computed point, one can expect that $\Delta(0)$ for the true---not variational---instanton will touch zero at some current $I_*\approx 0.9\,I_c$. One can further argue that this $I_*$ marks the point of the second-order transition between topologically nontrivial (vortex-like) and topologically trivial instantons. In such a scenario, the Boussinesq instanton at $I\to I_c$ continuously transforms into a vortex-antivortex pair, with vanishing $\bj(0)$ and $\Delta(0)$ at $I_\text{top}$ and further splitting of the emerging zero in $\Delta(\br)$ into a pair of zeros of positive and negative windings at $I<I_\text{top}$ (similar to the mechanism of topological transition in stretched graphene \cite{stretched-graphene-1,stretched-graphene-2}). The fact that our probe numerical solution can be extended to $\Delta(0)\approx 0.1\,\Delta_0$ presumably excludes an alternative  first-order scenario, when the vortex-antivortex and Boussinesq-like solutions coexist in some region, with their barriers being equal at $I_\text{top}$.

Our results derived for an infinite plane determine the dark-count rate in SNSPD detectors with wide superconducting strips \cite{SNSPD-wide-Korneeva,SNSPD-wide-Charaev}. If the strip's width $w\gg L_y$, the instanton is formed right at its edge and has the activation energy $\Delta F^\text{2D}/2$. In the opposite limit of narrow strips ($w\ll L_y$), the problem becomes 1D, with the barrier height given by the LA expression \eqref{F-LA}.
Qualitatively, our findings agree with the numerics presented in Ref.~\cite{Vodolazov2012} for strips of finite width. The same supercurrent dependence of $\Delta F$ with exponent 3/4 was
reported in Ref.\ \cite{OV2015}. However, their activation barrier is extensive in the strip width, $\Delta F_\text{OV}\sim(w/\xi)\Delta F^\text{2D}$, making that solution irrelevant to 2D phase-slip physics.

To conclude, we provide the first analytical calculation of the activation barrier for thermal phase slips in infinite current-carrying superconducting films at $I\to I_c$. In this limit, the saddle-point configuration is governed by the exactly integrable Boussinesq equation. The instanton is strongly anisotropic with $L_y\gg L_x\gg\xi$, leading to the $(1-I/I_c)^{3/4}$ scaling of the activation barrier. 
Based on numerical simulations, we conjecture a second-order topological transition between a vortex-free and a vortex-antivortex saddle-point configurations at $I_\text{top}\approx0.9\,I_c$.

We are grateful to I. V. Kolokolov, V. P. Mineev, and A. V. Semenov for stimulating discussions. This research was supported by the Russian Science Foundation under Grant No.\ 23-12-00297 and by the Foundation for the Advancement of Theoretical Physics and Mathematics ``BASIS''.


\end{document}